\documentclass[nofootinbib,a4paper,fleqn,aps,prd,10pt,superscriptaddress,reprint,showkeys,showpacs]{revtex4-1}

\usepackage{times}
\usepackage{amsfonts}
\usepackage{amsmath}
\usepackage{amssymb}
\usepackage{natbib}
\usepackage{graphicx}
\usepackage{enumitem}
\usepackage{xcolor}
\usepackage[outdir=./]{epstopdf}
\usepackage[normalem]{ulem}
\usepackage[colorlinks]{hyperref}
\usepackage{hyperref}
\hypersetup{%
    ,urlcolor=blue
    ,citecolor=blue
    ,linkcolor=blue
    }

\def\aap{A\&A}
\def\aj{AJ}
\def\apj{ApJ}
\def\apjl{ApJL}
\def\apjs{ApJS}
\def\apss{Astrophysics and Space Science}
\def\jcap{JCAP}
\def\mnras{MNRAS}

\def\physrep{Physics Reports}

\begin{document}

\title{Mass-Temperature relation in \texorpdfstring{$\Lambda$CDM}{LCDM} and modified gravity}

\author{Antonino~\surname{Del Popolo}}%
\affiliation{%
Dipartimento di Fisica e Astronomia, University of Catania, Viale Andrea Doria 6, 95125, Catania, Italy
}
\affiliation{%
INFN sezione di Catania, Via S. Sofia 64, I-95123 Catania, Italy
}
\email[Corresponding author.\\]{adelpopolo@oact.inaf.it}

\author{Francesco~\surname{Pace}}%
\affiliation{%
Jodrell Bank Centre for Astrophysics, School of Physics and Astronomy, The University of Manchester, Manchester, 
M13 9PL, United Kingdom
}
\email[]{francesco.pace@manchester.ac.uk}

\author{David~F.~\surname{Mota}}%
\affiliation{%
Institute of Theoretical Astrophysics, University of Oslo, P.O. Box 1029 Blindern, N-0315 Oslo, Norway 
}
\email[]{d.f.mota@astro.uio.no}

\label{firstpage}

\date{\today}

\begin{abstract}
We derive the mass-temperature relation using an improved top-hat model and a continuous formation model which takes 
into account the effects of the ordered angular momentum acquired through tidal-torque interaction between clusters, 
random angular momentum, dynamical friction, and modifications of the virial theorem to include an external pressure 
term usually neglected. We show that the mass-temperature relation differs from the classical self-similar behavior, 
$M \propto T^{3/2}$, and shows a break at $3--4$~keV, and a steepening with a decreasing cluster temperature. We then 
compare our mass-temperature relation with those obtained in the literature with $N$-body simulations for $f(R)$ and 
symmetron models. 
We find that the mass-temperature relation is not a good probe to test gravity theories beyond Einstein's general 
relativity, because the mass-temperature relation of the $\Lambda$CDM model is similar to that of the modified 
gravity theories.
\end{abstract}



\maketitle

\section{Introduction}
The wealth of astronomical observations available nowadays clearly shows either that our Universe contains more 
mass-energy than is seen or that the accepted theory of gravity, general relativity (GR), is somehow not correct, or 
both \cite{Bull2016}. 
The central assumption of the concordance $\Lambda$CDM model relies on gravity being correctly described by GR so 
that dark matter (DM), a nonbaryonic and nonrelativistic particle, and dark energy (DE), in the form of the 
cosmological constant $\Lambda$, constitute its dominant components \cite{DelPopolo2014}. 
Despite gravitational evidence for DM from galaxies \cite{Bertone2005}, cluster of galaxies \cite{Battistelli2016}, 
cosmic microwave background (CMB) anisotropies \cite{Bouchet2004}, cosmic shear \cite{Kilbinger2015}, structure 
formation \cite{DelPopolo2007}, and large-scale structure of the Universe \cite{Einasto2001}, decades of direct and 
indirect searches of those DM particles did not give any positive result \cite{Klasen2015}. 
In addition, the accelerated expansion of the universe modeled with $\Lambda$ \cite{Riess1998} raised the 
``cosmological constant fine-tuning problem", and the ``cosmic coincidence problem" 
\cite{Astashenok2012,Velten2014,Weinberg1989}.

The success of the $\Lambda$CDM model in describing the formation and evolution of the large-scale structures in the 
Universe at early and late times \cite{Spergel2003,Komatsu2011,DelPopolo2007} cannot hide the tensions at small 
\cite{Moore1999,deBlok2010,Ostriker2003,BoylanKolchin2011,DelPopolo2014a,DelPopolo2014d,DelPopolo2017a} and large 
scales \citep{Eriksen2004,Schwarz2004,Cruz2005,Copi2006,Macaulay2013,Planck2014_XVI,Raveri2016} precision data are 
currently revealing.

Small-scale problems \cite{DelPopolo2017a} have sprung two sets of attempts of solutions to save the $\Lambda$CDM 
paradigm: cosmological and astrophysical recipes. The first are based on either modifying the power spectrum on small 
scales \citep{Zentner2003} or altering the kinematic or dynamical gravitational behavior of the constituent DM 
particles. The latter, like supernovae (SN) feedback \cite{Brooks2013,Onorbe2015,DelPopolo2017a} and transfer of energy 
and angular momentum from baryon clumps to DM through dynamical friction 
\cite{ElZant2001,ElZant2004,DelPopolo2009,Nipoti2015,DelPopolo2016a}, rely on some ``heating" mechanism producing an 
expansion of the galaxy's DM component which reduces its inner density.

The previous issues seeded the push for several new modified gravity (MG) theories, to understand our Universe without 
DM \citep{Bekenstein2010} or at least to connect the accelerated expansion to some new features of gravity 
\citep{joyce2016}.

A first drive for MG came from fundamental problems in the hot big bang model (horizon, flatness and monopole 
problem solved within the inflationary paradigm \citep{Starobinsky1979,Guth1981}) and another one from galaxy rotation 
curves with solutions attempted within the modified Newtonian dynamics (MOND) \citep{Milgrom1983} and the ``modified 
gravity" (MOG) paradigm \citep{Moffat2006} and $f(R)$ theories \citep{DeFelice2010}.

Alternative proposals to explain the accelerated expansion of the Universe increased exponentially. 
Besides DM-like DE schemes \cite{ArmendarizPicon2001,Kamenshchik2001,dePutter2007,Durrer2008}, MG theories attempted 
to explain such acceleration as the manifestation of extra dimensions, or higher-order corrections effects, as in the 
Dvali-Gabadadze-Porrati model \cite{Dvali2000} and in $f(R)$ gravity. 
Nowadays, the catalog of MG theories includes many theories, among which we recall $f(R)$ \citep{DeFelice2010}, 
$f(T)$ \citep{Linder2010}, MOND and BIMOND \citep{Milgrom1983,Milgrom2014}, tensor-vector-scalar theory 
\citep{Bekenstein2004},  scalar-tensor-vector gravity theory (MOG) \citep{Moffat2006}, Gauss-Bonnet models
\citep{Zwiebach1985,Nojiri2005}, Lovelock models \citep{Lovelock1971}, Ho{\v r}ava-Lifshitz \citep{Horava2009}, 
Galileons \citep{Rodriguez2017}, and Horndeski \citep{Horndeski1974,Deffayet2010}. 
The freedom allowed to MG from observations reduces to modifications on large scales (typically Hubble scales), low 
accelerations ($a_0 \lesssim 10^{-8}~{\rm cm\,s}^{-2}$), or small curvatures (typically 
$R_{\Lambda} \simeq 1.2 \times 10^{-30} R_{\odot}$ \citep{Debono2016}). 
Some theories violate Birkhoff's theorem and this induces effects that should be disentangled wisely as they make 
local tests complex. Such local tests, using PPN-like parameters\footnote{The parametrized post-Newtonian (PPN) 
formalism is a tool expressing Einstein's equations in terms of the lowest-order deviations from Newton's law of 
gravitation.} \citep{Ni1972,Will1993,Bertotti2003,Will2014} and the GR condition on the two Newtonian potentials 
$\Phi=\Psi$, provide a smoking gun for MG, combining galaxy surveys ($\propto \Phi$), the integrated Sachs-Wolfe 
effect \cite{Dupe2011} in the CMB [$\propto \int dl (\dot \Phi+ \dot \Psi)$], and weak lensing 
[$\propto \int dl (\Phi+\Psi)$]. 
Real opportunities will come with future surveys: both from satellites 
(Euclid \citep{Euclid} and JDEM \citep{JDEM}) and ground-based (SKA \citep{SKA} and LSST \citep{LSST}). Another smoking 
gun should proceed from the best fitting of the CMB between DM and MG to constrain the parameters of the models 
\citep{Battye2018a,Battye2019}.

For MG theories not to alter the behavior of gravity at small scales (e.g., Solar System) and reproduce the 
observational measurements \cite{Dimopoulos2007,Bertotti2003},it is necessary to have some screening mechanism which
hides undesired effects on small scales \citep{Brax2012}. 
Following Ref.~\cite{Hammami2017}, we consider the case of the symmetron scalar-tensor theory \citep{Hu2007} and the 
chameleon $f(R)$ gravity \cite{Hinterbichler2010}.

Effects of MG can be probed with structure formation and verified by means of dark-matter-only $N$-body simulations 
\cite{Llinares2008,Zhao2011,Puchwein2013,Llinares2014,Gronke2014}. 
Nevertheless, hydrodynamical simulations are more suited from an observational point of view, as they provide 
observables, such as the halo profile, the turnaround  \cite{Bhattacharya2017,Lopes2018}, the splashback radius 
\cite{Adhikari2018}, and the mass-temperature relation (MTR) \cite{Hammami2017} which can be directly compared with 
observations. While the halo profile is usually studied in DM-only simulations and it is, as such, used for a variety 
of studies, the MTR can be accurately inferred only with hydrodynamic simulations, to avoid the necessary 
approximations introduced, for example, by using scaling relations. 
The MTR has been used to put constraints on MG theories. By means of hydrodynamic simulations, Ref.~\cite{Hammami2017} 
showed that the MTR obtained in MG theories is different from the expectations of GR.\footnote{In the literature, 
there is no explicit emphasis on what is exactly meant for mass. In general, when considering both numerical 
simulations and observations, the mass has to be the virial mass, as a result of the application of the virial theorem. 
This is more appropriately true for observations but less for $N$-body simulations, as the spherical overdensity 
procedure obtained to infer structures assumes a virial overdensity but does not automatically imply the virial 
theorem holding. Furthermore, the virial overdensity chosen will depend on which probe is considered (i.e., SZ effect 
or x-ray emission); therefore, the virial mass will be interpreted differently in different scenarios. We therefore 
prefer to just call it mass, having in mind it is related to the true virial mass of the object.}

In the present paper, we extended the results of Ref.~\citep{DelPopolo2002} to take into account the effects of 
dynamical friction and the cosmological constant and revisited the results of Ref.~\cite{Hammami2017} to show that the 
MTR is not a good probe to disentangle MG from GR. To this aim, we use a semianalytic model to show that in a 
$\Lambda$CDM model the MTR has a behavior similar to those obtained by Ref.~\cite{Hammami2017}, and this makes it 
impossible to disentangle between the MG results and those of GR.

The paper is organized as follows. Section~\ref{sect:MG} briefly presents the modified gravity models analyzed in this 
work, while Sec.~\ref{sect:model} describes the model used to derive the MTR relation in $\Lambda$CDM cosmologies. 
Section~\ref{sect:results} is devoted to the presentation and the discussion of our results. We conclude in 
Sec.~\ref{sect:conclusions}.

In this work, we use the following cosmological parameters: $h_0 = 0.7$, $\Omega_{\Lambda} = 0.727$, 
$\Omega_{\rm DM}= 0.227$, and $\Omega_{\rm b}= 0.046$. An overbar will indicate quantities evaluated at the background 
level.

\section{Modified gravity: models and simulations}\label{sect:MG}
In this section, we summarize the modified gravity theories used by Ref.~\citep{Hammami2017} that we compare our model 
to. These are scalar-tensor theories of gravity described by the action
\begin{eqnarray}\label{action}
 S & = & \int \mathrm{d}^4x\sqrt{-g} \left[\frac{1}{2}M_{\rm pl}^2R - \frac{1}{2}\partial^i\varphi\partial_i\varphi - 
                                           V(\varphi)\right] \nonumber\\
   & + & S_{\rm m}(\tilde{g}_{\mu\nu}, \varphi_i)\,,
\end{eqnarray}
where $g$ is the determinant of the metric tensor $g_{\mu\nu}$, $R$ the Ricci scalar, $M_{\rm pl}=1/\sqrt{8\pi G}$ the 
reduced Planck mass (in natural units where $\hbar=c=1$), and $\varphi$ and $V(\varphi)$ the scalar field and the 
self-interacting potential, respectively. 
Matter is described by the total matter action $S_{\rm m}$. The scalar field is conformally coupled to matter via 
$\tilde{g}_{\mu\nu}=A(\varphi)g_{\mu\nu}$, with $A(\varphi)$ the conformal factor.

The conformal coupling between matter and field gives rise to a fifth force of the form
\begin{equation}
 F_{\varphi} = -\frac{A^{\prime}({\varphi})}{A(\varphi)}\nabla\varphi\,,
\end{equation}
where a prime indicates the derivative with respect to the scalar field.

\subsection{Symmetron}
The screening mechanism of the symmetron model \citep{Hinterbichler2010} produces a strong coupling between matter and 
the extra field in low-density regions, while in high-density regions the scalar degree of freedom decouples from 
matter.

For this mechanism to work, one requires, around $\varphi = 0$, a coupling of the form
\begin{equation}
 A(\varphi) = 1 + \frac{1}{2}\left(\frac{\varphi}{M}\right)^2
\end{equation}
and a potential
\begin{equation}
 V(\varphi) = V_0 - \frac{1}{2}\mu^2\varphi^2 + \frac{1}{4}\lambda\varphi^4\,,
\end{equation}
where $M$ and $\mu$ are mass scales and $\lambda$ a dimensionless parameter. 

The free parameters can be recast in terms of the strength of the scalar field, $\beta$, the expansion factor at the 
symmetry breaking time, $a_{\rm SSB}$, and the range of the fifth force, $\lambda_0$. The fifth force then reads
\begin{equation}
 F_{\varphi} = -\frac{\varphi}{M^2}\nabla \varphi
             = 6\Omega_{\rm m} H^2_0 \frac{\beta^2\lambda_0^2}{a^3_{\rm SSB}}\tilde{\varphi}\nabla\tilde{\varphi}\,,
\end{equation}
where the quantities with tilde are in the supercomoving coordinates \citep{Martel1998}.

\subsection{\texorpdfstring{$f(R)$}{fR} gravity}\label{sect:fR-gravity} 
The $f(R)$-gravity models are theories in which the Ricci scalar in the Einstein-Hilbert action is substituted by a 
function of the same quantity, and it is described by the following action
\begin{equation}\label{husawicki}
 S = \frac{1}{2}M_{\rm pl}^2\int\mathrm{d}^4x\sqrt{-g}\left[R + f(R)\right]\,.
\end{equation}
When $f(R)=-2\Lambda$, the $\Lambda$CDM model is recovered.

Typical of these theories is the chameleon screening mechanism, characterized by a local density dependence of the 
scalar field mass. In high-density environments, the scalar degree of freedom is very short ranged, and the opposite 
happens in low-density fields, where deviations from GR are maximized.

Reference \citep{Hammami2017} used the Hu-Sawicki \citep{Hu2007} model, whose functional form is
\begin{equation}
 f(R) = -m^{2}\frac{c_1(R/m^2)^n}{1 + c_2(R/m^2)^n}\,,
\end{equation}
where the free parameter $m^2 = H_0^2\Omega_{\rm m,0}$ has dimensions of mass squared and $n>0$. The two additional 
constants $c_1$ and $c_2$ can be determined by requiring that in the large curvature regime ($R/m^2\gg 1$),  
$f(R)\approx -2\Lambda$
\begin{equation}
 \frac{c_1}{c_2} \approx 6\frac{\Omega_{\Lambda,0}}{\Omega_{\rm m,0}}\,.
\end{equation}

The strength of gravity modifications is encoded in the value of $f_R=df/dR$ today
\begin{equation}
 f_{R0} = -n\frac{c_1}{c_2^2}\left[\frac{\Omega_{\Lambda,0}}{3(\Omega_{\rm m,0} + 4\Omega_{\Lambda,0})}\right]^{n+1}\,.
\end{equation}
The range of the scalar degree of freedom is $\lambda_0 \propto \sqrt{1/f_{R0}}$.

To derive the expression of the fifth force for $f(R)$ models, it is useful to transform them into scalar-tensor 
theories using the conformal transformation $A(\varphi) = \exp{(-\beta\varphi/M_{\rm pl}})$, where $\beta=\sqrt{6}/6$. 
We then find
\begin{align}
 F_{\varphi} = -\frac{a^2\beta}{M_{\rm pl}}\nabla\varphi\,,
\end{align}
with $a$ the scale factor.

\subsection{Simulations}
In order to get the MTR for $f(R)$ and symmetron models, Ref.~\cite{Hammami2017} modified the ISIS code 
\citep{Llinares2014} and ran two sets of simulations, one for $f(R)$-gravity models and another one for the symmetron 
models, both containing $256^3$ DM particles. The box size and background cosmology were different for the two models, 
due to consistency with previous works of the authors \citep{Hammami2015}. In the case of the $f(R)$ gravity 
(symmetron) the DM particle mass was $3\times10^{10} M_{\odot}/h$ ($8.32\times10^{10} M_{\odot}/h$), $\Omega_{\Lambda} = 
0.727$, $\Omega_{\rm CDM} = 0.227$ and $\Omega_{\rm b} = 0.045$ ($\Omega_{\Lambda} = 0.65$, $\Omega_{\rm CDM} = 0.3$ 
and $\Omega_{\rm b} = 0.05$), and the box size $200~{\rm Mpc}~h^{-1}$ ($256~{\rm Mpc}~h^{-1}$), with $h = 0.7$ 
($h = 0.65$).

Because of the different parameters for $f(R)$ and symmetron models, the background $\Lambda$CDM model of the two 
models is different. Table~2 in Ref.~\citep{Hammami2017} summarizes the parameters employed.

\section{The Model}\label{sect:model}
In the next sections, we will discuss how the top-hat model (THM) can be improved, and how the MTR is calculated. 
We show two different models, the ``late-formation approximation" (see the following) and a model in which structures 
form continuously.

\subsection{Improvements to the top-hat model}
Using scaling arguments, one can show that there exists a relation between the x-ray mass of clusters and their 
temperature $T_{\rm X}$. The mass in the virial radius can be written as 
$M(\Delta_{\rm vir}) \propto T_{\rm X}^{3/2} \rho_{\rm c}^{-1/2} \Delta_{\rm vir}^{-1/2}$, where $\rho_{\rm c}$ is the 
critical density, and $\Delta_{\rm vir}$ the density contrast of a spherical top-hat perturbation after collapse and 
virialization.

The previous relation shows a correlation between the mass and temperature, but this result can be highly improved. One 
possibility is to improve the THM, taking into account the angular momentum acquired by the interaction with 
neighboring protostructures, dynamical friction, and a modified version of the virial theorem, including a surface 
pressure term \citep{Voit1998,Voit2000,Afshordi2002,DelPopolo2005} due to the fact that at the virial radius 
$r_{\rm vir}$ the density is different from zero, as done in \citep{DelPopolo1999}.

A further improvement can be obtained by taking into account that clusters form in a quasicontinuous way. To this aim, 
one substitutes the top-hat cluster formation model by a model of cluster formation from spherically symmetric 
perturbations with negative radial density gradients. The merging-halo formalism of Ref.~\cite{Lacey1993} is used to 
take into account the gradual way clusters form.

To start with, we consider some gravitationally growing mass concentration collecting into a potential well. Let 
${\rm d}P=f(L,r,v_r,t){\rm d}L{\rm d}v_r{\rm d}r$ be the probability that a particle, having angular momentum 
$L=r v_{\theta}$, is located at $[r,r+{\rm d}r]$, with velocity ($v_r={\dot r}$) $[v_r,v_r+{\rm d}v_r]$, and 
angular momentum $[L,L+{\rm d}L]$. The term $L$ takes into account ordered angular momentum generated by tidal 
torques and random angular momentum (see Appendix C.2 in Ref.~\citep{DelPopolo2009}). The radial acceleration of the 
particle \citep{Peebles1993,Bartlett1993,Lahav1991,DelPopolo1998,DelPopolo1999} is
\begin{equation}\label{eq:coll}
 \frac{{\rm d}v_r}{{\rm d}t} = -\frac{GM}{r^2} + \frac{L^2(r)}{M^{2}r^3} + \frac{\Lambda}{3}r -
 \eta\frac{{\rm d}r}{{\rm d}t}\,,
\end{equation}
with $\Lambda$ being the cosmological constant and $\eta$ the dynamical friction coefficient. 
The previous equation can be obtained via Liouville's theorem \citep{DelPopolo1999}. 
The last term, the dynamical friction force per unit mass, is more explicitly given in Ref.~\citep{DelPopolo2009}
[Appendix D, Eq.~(D5)]. 
A similar equation (excluding the dynamical friction term) was obtained by several authors (see, e.g., 
\citep{Fosalba1998a,Engineer2000,DelPopolo2013b}) and generalized to smooth dark energy models in 
Ref.~\cite{Pace2018}.

In the framework of general relativity, Refs.~\citep{Pace2010,Pace2017a} derived the nonlinear evolution equation of 
the overdensity $\delta=\delta\rho_{\rm m}/\bar{\rho}_{\rm m}$ of nonrelativistic matter
\begin{equation}\label{eqn:nleq1}
 \begin{split}
  \ddot{\delta}+2H \dot{\delta}-\frac{4}{3}\frac{\dot{\delta}^2}{1+\delta}-
  4\pi G\bar{\rho}_{\rm m} \delta(1+\delta)-\\
  (1+\delta)(\sigma^2-\omega^2) & = 0\,.
 \end{split}
\end{equation}
Recalling that $\delta=\frac{2GM_{\rm m}}{\Omega_{m,0} H^2_0}(a/R)^3-1$, where $R$ is the effective perturbation 
radius and $a$ the scale factor, substituting into Eq.~(\ref{eqn:nleq1}) one gets \citep{Pace2018}
\begin{equation}
 \ddot{R} = -\frac{GM_{\rm m}}{R^2} - \frac{GM_{\rm de}}{R^2}(1+3w_{\rm de})-\frac{\sigma^2-\omega^2}{3}R\,,
\end{equation}
where $M_{\rm m}$ and $M_{\rm de}$ are the matter mass content of the perturbation and the mass of the dark energy 
component, respectively. The previous equations can be generalized to account for the presence of dynamical friction 
using Eckart's formalism \citep{Barbosa2015}. The standard Friedmann equation is now augmented with a fluid describing 
the contribution of the viscosity
\begin{equation}
 \left(\frac{\dot a}{a}\right)^2 = H^2 = \frac{8\pi G}{3}(\bar{\rho}_{\rm v} + \bar{\rho}_{\rm m} + 
                                                          \bar{\rho}_{\Lambda})\,,
\end{equation}
where $\bar{\rho}_{\Lambda}$ is the energy density of the cosmological constant, 
$\bar{\rho}_{\rm m} = \bar{\rho}_{\rm m,0}a^{-3}$ the matter component and 
$\dot{\bar{\rho}}_{\rm v}+3H\bar{\rho}_{\rm v} = 3H^2\xi_0\bar{\rho}_{\rm v}^\nu$ the viscous component, with 
$\xi_0$ the bulk viscosity coefficient. The bulk viscosity is expressed as $\xi=\xi_0\bar{\rho}_{\rm v}^{\nu}$, where 
$\nu$ is a real constant.

Integrating Eq.~(\ref{eq:coll}) with respect to $r$, we have:
\begin{equation}\label{eq:coll1}
 \frac{1}{2}\left(\frac{{\rm d}r}{{\rm d}t}\right)^{2} = \frac{GM}{r} + 
 \int_0^r \frac{L^{2}}{M^{2}r^{3}}\mathrm{d}r + \frac{\Lambda}{6}r^{2} - \int_0^r 
 \eta\frac{{\rm d}r}{{\rm d}t} + \epsilon\,.
\end{equation}
The specific binding energy of the shell, $\epsilon$, can be obtained from the turnaround condition 
$\frac{{\rm d}r}{{\rm d}t}=0$.

One can obtain the MTR combining energy conservation, the virial theorem, using Eq.~(\ref{eq:coll1}) and the 
connection between kinetic energy $K$ and the temperature \citep{Afshordi2002}:
\begin{equation}\label{eq:conn}
 \langle K \rangle = \frac{3\tilde{\beta} M k_{\rm B} T}{2 \mu m_{\rm p}}\,,
\end{equation}
where $\mu=0.59$ is the mean molecular weight, $k_{\rm B}$ is the Boltzmann constant, $m_{\rm p}$ the proton mass, 
$\tilde{\beta} = \beta[1+f(1/\beta-1)\Omega_{\rm b,0}/\Omega_{\rm m,0}]$, $\Omega_{\rm b,0}$ ($\Omega_{\rm m,0}$) is 
the baryonic (total) matter density parameter today, $f$ is the fraction of the baryonic matter in the hot gas, and the 
parameter $\beta=\frac{\mu m_{\rm p}\sigma_{\rm v}^2}{k_{\rm B}T}$, $\sigma_{\rm v}$ being the ratio of the 
mass-weighted mean velocity dispersion of the dark matter particles.

Using the virial theorem, we have \citep{Landau1966,Lahav1991,DelPopolo1999}
\begin{equation}\label{eq:virial}
 \langle K \rangle = \frac{3\tilde{\beta} M k_{\rm B} T}{2 \mu m_{\rm p}} = 
                    -\frac{1}{2} \langle U_{\rm G} \rangle - \langle U_{\rm L} \rangle + \langle U_{\Lambda}\rangle 
                    + \langle U_{\eta} \rangle\,.
\end{equation}
The brackets indicate time average (see \citep{Bartlett1993}). The four terms represent the energy related to the 
gravitational potential, the angular momentum, the cosmological constant, and the dynamical friction, respectively.

Equation~(\ref{eq:virial}) does not take into account the surface pressure term we spoke about, though. Assuming 
\citep{Afshordi2002}
\begin{equation}\label{eqn:pressure}
 \langle K \rangle + \langle E \rangle = 3 P_{\rm ext} V = -\nu U\,,
\end{equation}
with $V$ the volume of the outer boundary of the virialized region, $P_{\rm ext}$ the pressure on the boundary, $\nu$ 
a constant and $U$ the total potential (see \citep{Afshordi2002}), Eq.~(\ref{eq:virial}) reads now
\begin{equation}\label{eq:virial1}
 \langle K \rangle = (1+\nu)\left(-\tfrac{1}{2} \langle U_{\rm G}\rangle - \langle U_{\rm L} \rangle + 
                                  \langle U_{\Lambda} \rangle + \langle U_{\eta} \rangle \right)\,.
\end{equation}
In other words, the averaged kinetic energy differs by a factor $1+\nu$ from before.

In order to estimate the effect of the boundary pressure on the virial theorem, we consider an isothermal velocity 
dispersion ($\sigma_{\rm 1D} = \mathrm{const}$), and then $P=\rho_{\rm vir}\sigma_{\rm 1D}^2$, for which we 
have \citep{Voit2000}
\begin{equation}
 \langle K \rangle = \frac{\bar{\rho}_{\rm m,vir}}{2\rho(r_{\rm vir}) - \bar{\rho}_{\rm m,vir}} \langle E \rangle\,,
\end{equation}
where $\overline{\rho}$ is the mean density within the virial radius. 
If the local density is negligible at $r_{\rm vir}$, the confining pressure is zero and 
$\langle K \rangle = - \langle E \rangle$. For a Navarro-Frenk-White profile and a typical cluster value of the 
concentration parameter $c \simeq 5$, we have $|\langle K \rangle/\langle E \rangle| \simeq 2$. 
References \citep{Shapiro1999,Iliev2001} studied in detail the effect of the quoted boundary pressure, finding that 
it changes significantly the final object. More in detail, it is found that the virial temperature is affected (larger 
than a uniform sphere but smaller than a truncated singular approximation sphere) and the extrapolated linear 
overdensity contrast $\delta_{\rm c}$ is slightly smaller, implying an earlier collapse.

We now use energy conservation in the form (see \citep{Lahav1991,DelPopolo1999})
\begin{eqnarray}\label{eq:energy}
\langle E \rangle & = & \langle K \rangle + \langle U_{\rm G} \rangle + \langle U_{\rm \Lambda} \rangle + 
                        \langle U_{\rm L} \rangle+ \langle U_{\eta} \rangle\ \\
                  & = & U_{\rm G,ta}+U_{\Lambda,{\rm ta}}+U_{\rm L,ta}+U_{\rm \eta,ta}\,,\nonumber
\end{eqnarray}
where the subscript ``${\rm ta}$" stands for turnaround.

Combining Eqs.~(\ref{eq:virial1}) and (\ref{eq:energy}), solving for $\langle K \rangle$ and recalling 
Eq.~(\ref{eq:conn}), we obtain
\begin{eqnarray}
 \frac{k_{\rm B}T}{{\rm keV}} & = & 1.58\left(\nu +1\right) \frac{\mu}{\beta}\frac{1}{\psi\xi}
                  \Omega_{\rm m,0}^{1/3}\left(\frac{M}{10^{15}M_{\odot}h^{-1}}\right)^{2/3}(1+z_{\rm ta}) \nonumber \\
             & &  \times \left[1+\left(\frac{32\pi}{3}\right)^{2/3} \psi\xi \bar{\rho}_{\rm m,ta}^{2/3}
                  \frac{1}{H_{0}^{2}\Omega_{\rm m,0}M^{8/3}(1+z_{\rm ta})}\right. \nonumber\\
             & &  \phantom{\times \left[\right.}
                  \times \int_0^{r_{\rm eff}} \frac{L^{2}}{r^{3}}{\rm d}r - 
                  \frac{2}{3}\frac{\Lambda}{\Omega_{\rm m,0}H_0^{2}(1+z_{\rm ta})^3}\left(\psi\xi\right)^3 \nonumber\\
             & &  \phantom{\times \left[\right.}
                  -\frac{2^{10/3}}{3^{2/3}} \pi^{2/3} \left(\frac{\psi\xi}{\Omega_{\rm m,0} H_0^2}\right)
                  \left(\frac{\rho_{\rm m,0}}{M}\right)^{2/3}
                  \frac{1}{1+z_{\rm ta}} \times\nonumber\\
             & &  \phantom{\times \left[\right.}\left.\int \eta \frac{{\rm d}r}{{\rm d}t}{\rm d}r\right]\,, 
                  \label{eq:tem} 
\end{eqnarray}
where $r_{\rm eff}=\psi r_{\rm ta}=\psi \xi \left(\frac{2GM}{\Omega_{\rm m,0}H_{0}^{2}}\right)^{1/3}$, $r_{\rm ta}$ is 
the radius at the turnaround epoch $z_{\rm ta}$, $\Omega_{\rm m,0}=\frac{8\pi G \bar{\rho}_{\rm m,0}}{3 H_0^2}$, 
$M=4 \pi \bar{\rho}_{\rm m,0} x^{3}_{1}/3$, and $ \xi=r_{\rm ta}/x_{1}$.

The product $\psi\xi$, using the definitions of $\psi$, $\xi$, and $M$ can be written as [see also Eq.~(\ref{eq:zdep}) 
and Ref.~\citep{Lilje1992}]
\begin{equation}\label{eq:psixi}
 \psi\xi = \frac{r_{\rm eff}}{r_{\rm ta}} \frac{r_{\rm ta}}{x_1}=\frac{r_{\rm eff}}{r_{\rm ta}}
           \left(\frac{\bar{\rho}_{\rm m,0}}{\rho_{\rm ta}}\right)^{1/3} (1+z_{\rm ta})^{-1} \,,
\end{equation}
where $\rho_{\rm ta}$ is the average density inside the perturbation at the turnaround.

Equation~(\ref{eq:tem}) can be also equivalently written, by using the notation of Ref.~\citep{Lilje1992}, in terms of 
$r_{\rm vir}$:
\begin{eqnarray}\label{eq:fin}
 \frac{k_{\rm B}T}{\rm keV} & = & 0.94\left(\nu+1\right)\frac{\mu}{\beta}\left(\frac{r_{\rm ta}}{r_{\rm vir}}\right)
                  \left(\frac{\rho_{\rm ta}}{\bar{\rho}_{\rm m,ta}}\right)^{1/3}\Omega_{\rm m,0}^{1/3}\nonumber\\
              &&  \times\left(\frac{M}{10^{15}M_{\odot}h^{-1}}\right)^{2/3}(1+z_{\rm ta})\nonumber\\ 
              &&  \times \left[1+\frac{15r_{\rm vir}\bar{\rho}_{\rm m,ta}}{\pi^{2}H_0^{2}\Omega_{\rm m,0}
                  \rho_{\rm ta}^{3}r_{\rm ta}^{9}(1+z_{\rm ta})}\int_0^{r_{\rm vir}} \frac{L^2 \mathrm{d}r}{r^3}
                  \right.\nonumber\\
              && -\frac{2}{3}\frac{\Lambda}{H_0^{2}\Omega_{m,0}}\left(\frac{r_{\rm vir}}{r_{\rm ta}}\right)^{3}
                  \left(\frac{\bar{\rho}_{\rm m,ta}}{\rho_{\rm ta}}\right) \frac{1}{(1+z_{\rm ta})^3}\nonumber\\
              && -\frac{6^{1/3}}{\pi^{1/3}}r_{\rm vir}r_{\rm ta}
                  \left(\frac{\bar{\rho}_{\rm m,ta}}{\rho_{\rm ta}}\right)^{1/3}
                  \left(\frac{\bar{\rho}_{\rm m,0}}{M} \right)^{2/3}\frac{1}{1+z_{\rm ta}}\times\nonumber\\
              &&  \left.\frac{\lambda_0}{1-\mu(\delta)}\right]\,,
\end{eqnarray}
where $\Omega_{\Lambda}=\frac{\Lambda}{3 H_0^2}=1-\Omega_{\rm m,0}$. 
In Eq.~(\ref{eq:fin}), we integrated the term containing the dynamical friction; $\lambda_0$ and $\mu(\delta)$ are 
given in Ref.~\citep{Colafrancesco1995}.  

The value of $r_{\rm eff}$, as shown in Refs.~\citep{Lahav1991,DelPopolo2002}, is given by the solution of the cubic 
equation:
\begin{eqnarray}
  1 & - & \nu+\left(\xi\psi\right)^{3}\left(\nu+2\right)\zeta - \psi \left(2+\zeta\xi^3\right) \nonumber\\
    & - & \frac{27}{32}\frac{\xi^{9}\psi}{\rho_{\rm ta}^{3}\pi^{3}Gr_{\rm ta}^{8}}
          \left[\nu\int_{0}^{r_{\rm eff}}\frac{L^{2}}{r^{3}}\mathrm{d}r + 
          \int_{0}^{r_{\rm ta}}\frac{L^{2}}{r^{3}}\mathrm{d}r \right.\nonumber\\
    & - & \frac{16\pi^2}{9}(2+\nu)\rho_{\rm ta}^2 r_{\rm ta}^6 \times \nonumber\\
    & &   \left.\left(
          \int_{0}^{r_{\rm eff}}\eta\frac{\mathrm{d}r}{\mathrm{d}t}\mathrm{d}r - 
          \frac{1}{2+\nu}\int_{0}^{r_{\rm ta}}\eta\frac{\mathrm{d}r}{\mathrm{d}t}\mathrm{d}r\right)\right]=0\,,
\end{eqnarray}
where 
\begin{equation}\label{eq:zdep}
 \zeta = \frac{\Lambda}{4 \pi G \rho_{\rm ta}} = \frac{\Lambda r_{\rm ta}^3}{3 G M} = 
         \frac{2\Omega_{\Lambda,0}}{\Omega_{m,0}}
         \frac{\bar{\rho}_{\rm m,ta}}{\rho_{\rm ta}}(1+z_{\rm ta})^{-3}\,.
\end{equation}
The parameter $\nu$, as shown by Ref.~\cite{Afshordi2002} [Eq.~(47)], depends on the concentration parameter and the 
density profile. We fixed it as $\frac{\nu+1}{\nu-1} \simeq 2$ \cite{Voit2000,Afshordi2002}, for a typical value of 
the cluster concentration parameter, $c \simeq 5$.

\subsection{Revisiting the continuous formation model}
The approximation in which we found the MTR is known as the late-formation approximation and assumes that perturbation 
clusters form from having a top-hat density profile and that the redshift of observation, $z_{\rm obs}$, is equal to 
that of formation, $z_{\rm f}$. The quoted approximation is good in the case $\Omega_{\rm m,0} \simeq 1$, where 
cluster formation is fast, and at all redshifts $z_{\rm obs} \simeq z_{\rm f}$. For the actual value of 
$\Omega_{\rm m,0}$, one needs to take into account the difference between $z_{\rm obs}$ and $z_{\rm f}$. 
Moreover, as shown by Ref.~\citep{Voit2000}, continuous accretion is needed to get the correct normalization of the MTR 
and its time evolution.

In order to improve the THM, one can take into account the formation redshift \citep{Kitayama1996,Viana1996} or the 
THM can be replaced by a model in which clusters form from spherically symmetric perturbations 
\citep{Voit2000,Voit1998}, combined with the merging-halo formalism of Ref.~\cite{Lacey1993}. In this way, one moves 
from a model in which clusters form instantaneously to one in which they form gradually.

Integrating Eq.~(\ref{eq:coll1}), one gets
\begin{equation}\label{eq:tmppp}
 t = \int\frac{\mathrm{d}r}{\sqrt{2\left[\epsilon + \frac{GM}r+\int_{r_{\rm i}}^{r} 
     \frac{L^2}{M^2r^3}\mathrm{d}r+\frac{\Lambda}{6}r^2\right] -\int \eta\frac{{\rm d}r}{{\rm d}t}{\rm d}r}}\,.
\end{equation}

Following Ref.~\citep{Voit2000}, we may write the specific energy of infalling matter as
\begin{equation}
 \epsilon_l = -\frac{1}{2}\left(\frac{2\pi GM}{t_\Omega}\right)^{2/3}
               \left[\left(\frac{M_0}{M}\right)^{5/(3m)}-1\right]g(M)\,,
\end{equation}
where $t_{\Omega}=\pi\Omega_{\rm m,0}/[H_0(1-\Omega_{\rm m,0})^{3/2}]$, $M_0$ is a fiducial mass, $m$ is a constant 
specifying how the mass variance evolves as a function of $M$ and the function $g(M)$ reads
\begin{equation}
 g(M) = 1+\frac{F}{x-1} +\frac{\lambda_0}{1-\mu(\delta)}+\frac{\Lambda}{3 H^2_0 \Omega_{\rm m,0}} \xi^3\,,
\end{equation}
where $x=1+(t_{\Omega}/t)^{2/3}$ is connected to the mass by $M=M_0 x^{-{3m/5}}$, $M_0$ is given in \citep{Voit2000}, 
and
\begin{equation}
 F = \frac{2^{7/3}\pi^{2/3}\bar{\rho}_{\rm m,0}^{2/3}}{3^{2/3}H^2_0\Omega_{\rm m,0}M^{8/3}}
     \int_{r_{\rm i}}^{r_{r_{\rm ta}}} \frac{L^2}{r^3}\mathrm{d}r\,.
\end{equation}
In order to calculate the kinetic energy $E$, we integrate $\epsilon_l$ with respect to the mass \citep{Voit2000} to 
get $-\int \epsilon_l dM=E/M$. Finally, we have
\begin{equation}
 k_{\rm B}T = \frac{4}{3}\tilde{a}\frac{\mu m_{\rm p}}{2\beta}\frac{E}{M}\,,
\end{equation}
where $\tilde{a} = \frac{\bar{\rho}_{\rm m,vir}}{2\rho(r_{\rm vir}) - \bar{\rho}_{\rm m,vir}}$ is the ratio between 
kinetic and total energy \citep{Voit2000} and $\bar{\rho}_{\rm m,vir}$ the mean density within the virial radius. 
Calculating $E/M$, we obtain
\begin{eqnarray}\label{eq:kT}
\frac{k_{\rm B}T}{\rm keV} & = & \frac{2}{5}\tilde{a}\frac{\mu m_{\rm p}}{2\beta} 
                                 \frac{m}{m-1}\left(\frac{2\pi G}{t_\Omega}\right)^{2/3}M^{2/3}
                                 \times\nonumber\\
                           & &   \left[\frac{1}{m} + \left(\frac{t_\Omega}{t}\right)^{2/3} + 
                                 \frac{K(m,x)}{(M/M_0)^{8/3}} + \frac{\lambda_0}{1-\mu(\delta)} \right.\nonumber\\
                           & &   \left.\quad +\frac{\Lambda\xi^3}{3H^2_0\Omega_{\rm m,0}}
\right]\,,
\end{eqnarray}
where
\begin{eqnarray}
 K(m,x) & = & (m-1)F x {\rm LerchPhi}(x,1,3m/5+1) - \nonumber \\
        & & (m-1) F {\rm LerchPhi}(x,1,3m/5)\,,
\end{eqnarray}
and {\rm LerchPhi} is a function defined as follows\footnote{This definition is valid for $|z| < 1$. By analytic 
continuation, it is extended to the whole complex $z$ plane for each value of $a$.}:
\begin{equation}
 {\rm LerchPhi}(z,a,v) = \sum_{n=0}^{\infty}\frac{z^n}{(v+n)^a}\,.
\end{equation}

Following Ref.~\citep{Voit2000} to get the normalization, Eq.~(\ref{eq:kT}) can be written as \citep{DelPopolo1999}
\begin{equation}\label{eq:kT1}
 k_{\rm B}T \simeq 8~{\rm keV} \left(\frac{M}{10^{15}h^{-1}M_{\odot}}\right)^{2/3}\frac{m(M)}{n(M)}\,.
\end{equation}
The functions $m(M)$ and $n(M)$ are defined as
\begin{eqnarray}
 m(M) & = & \frac{1}{m} + \left(\frac{t_\Omega}t\right)^{2/3} + \frac{K(m,x)}{(M/M_0)^{8/3}} + \nonumber\\
      &   & \frac{\lambda_0}{1-\mu(\delta)} + \frac{\Lambda\xi^3}{3H^2_0\Omega_{\rm m,0}} \,,\\
 n(M) & = & \frac{1}{m} + \left(\frac{t_{\Omega}}{t_{0}}\right)^{2/3} + K_0(m,x)\,,
\end{eqnarray}
where $K_0(m,x)$ indicates that $K(m,x)$ must be calculated assuming $t=t_0$.

When compared to Eq.~(17) of Ref.~\citep{Voit2000}, Eq.~(\ref{eq:kT1}) shows an additional mass-dependent term. This 
means that, as in the case of the top-hat model, the MTR is no longer self-similar, showing a break at the low-mass end 
(see next section).

Besides Refs.~\citep{Voit1998,Voit2000}, Ref.~\citep{Afshordi2002} found a MTR and its scatter. Their result concerning 
the MTR and the scatter is in agreement with the result we found here. In this case,
\begin{equation}
 k_{\rm B}T = 6.62~{\rm keV}~Q \left(\frac{M}{10^{15}~h^{-1}M_{\odot}}\right)^{2/3}\,,
\end{equation}
where
\begin{equation} \label{eq:afsh}
 Q = \frac{1+\nu}{1-\nu} \frac{B}{A(Ht)^{2/3}}
\end{equation}
and where $B/A$ is a constant [see the discussion after Eq.~(25) in Ref.~\citep{Afshordi2002}] and $\nu$ was defined in 
Eq.~(\ref{eqn:pressure}).

\section{Results and discussion}\label{sect:results}

\begin{figure*}[!ht]
 \centering
 \includegraphics[width=18cm,angle=0]{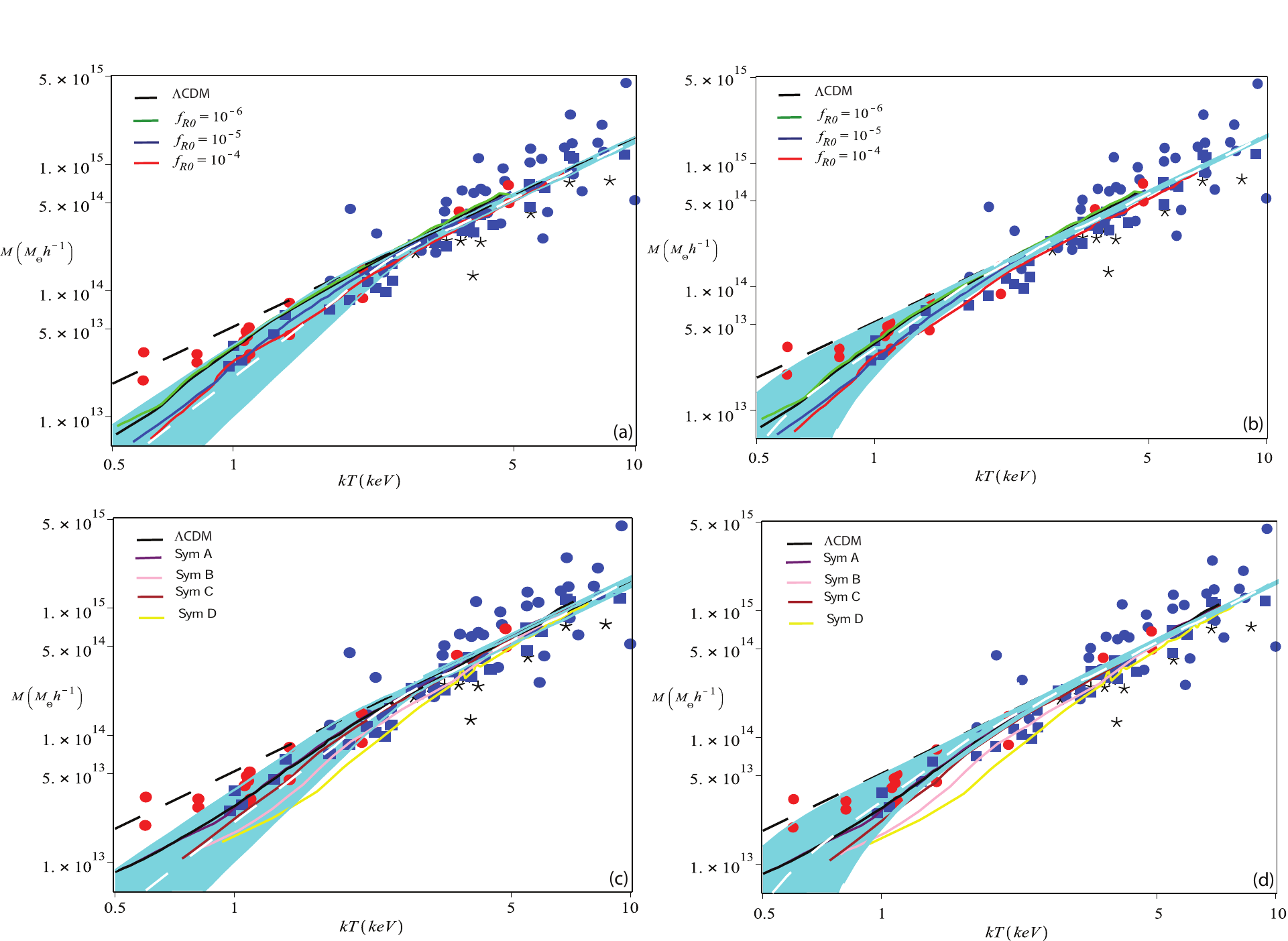}
 \caption[justified]{The MTR for $f(R)$ (top panels) and symmetron models (bottom panels). 
 In all the panels, the black line shows the $\Lambda$CDM model, and the dashed black line shows the MTR 
 $\propto T^{3/2}$ as obtained from scaling relations, while the stacked galaxy clusters are depicted with red and blue 
 circles, blue squares, and black stars. In panel (a) (top left) and in (b) (top right), the cyan region shows the 68\% 
 confidence level region, obtained using the continuous formation model [Eq.~(\ref{eq:kT1})] and the model by 
 Ref.~\citep{Afshordi2002} [Eq.~(\ref{eq:afsh})], respectively. The white dashed line is the average value. The red, 
 blue, and green lines represent the $f(R)$ model with three different normalizations. Panels (c) (bottom left) and (d) 
 (bottom right) are the equivalent of (a) and (b) for the symmetron models.}
 \label{fig:comparison}
\end{figure*}

\begin{figure*}[!ht]
 \centering
 \includegraphics[width=18cm,angle=0]{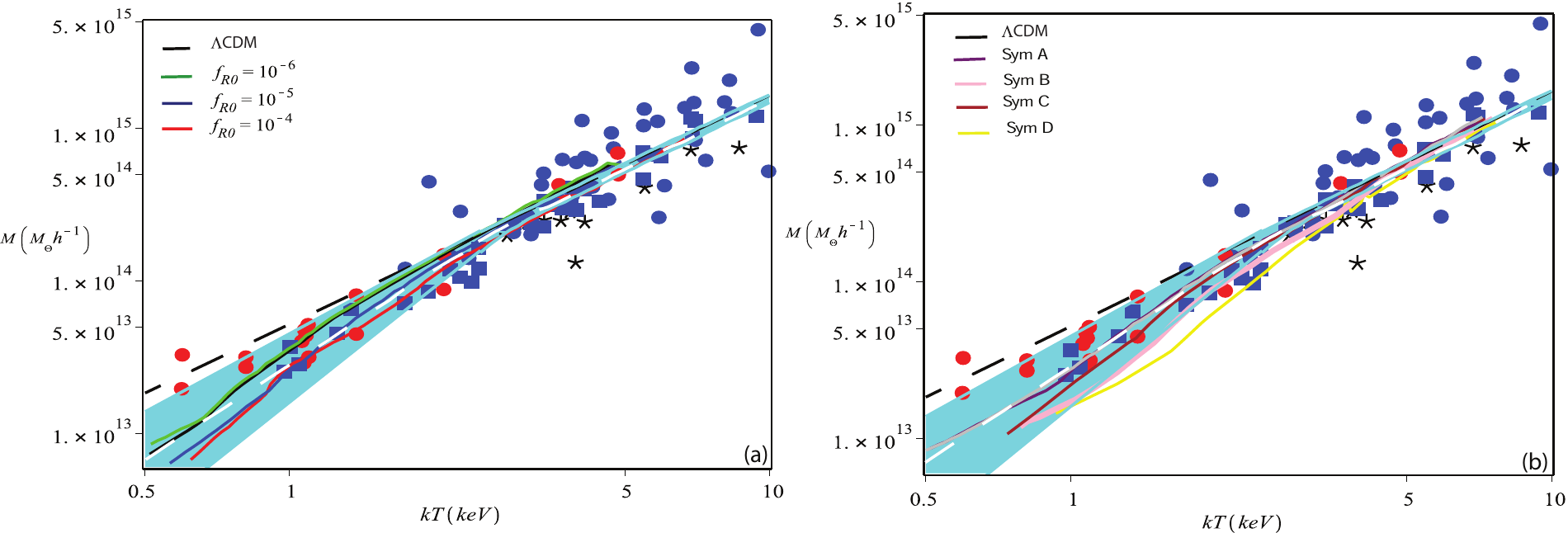}
 \caption[justified]{The MTR for $f(R)$ (left panel) and symmetron models (right panel). Lines and symbols represent 
 the same quantities as in Fig.~\ref{fig:comparison}, but now the cyan region is the 68\% confidence level region, 
 obtained by means of the improved top-hat model [Eq.~(\ref{eq:fin})].}
 \label{fig:comparison1}
\end{figure*}

In Fig.~\ref{fig:comparison}, we show the results of the comparison between our continuous formation model 
[Eq.~(\ref{eq:kT1})] and the model by Ref.~\citep{Afshordi2002} with that of \cite{Hammami2017} for $f(R)$ and 
symmetron models. For $f(R)$ models, we consider $n=1$ and $|f_{R0}|=10^{-4}, 10^{-5}, 10^{-6}$, while for the 
symmetron model $(\beta, a_{\rm SSB}, \lambda_0)=(1.0, 0.5, 1.0)$ for Sym A, (1.0, 0.33, 1.0) for Sym B, 
(2.0, 0.5, 1.0) for Sym C, and (1.0, 0.25, 1.0) for Sym D.

In all the panels, the black straight dashed line represents the classical MTR self-similar behavior and the black 
solid line the $\Lambda$CDM model obtained in the simulations of Ref.~\cite{Hammami2017} and for the specific modified 
gravity models we refer to the caption of Fig.~\ref{fig:comparison}. 
Observational data are represented by points. Red circles come from Ref.~\citep{Dai2007}, while blue points are from 
Ref.~\citep{Horner1999}. Stars are from Ref.~\citep{Horner1999} and represent data using spatially resolved 
observations.

Fig.~\ref{fig:comparison}(a) (top left panel) compares the result of our continuous formation model for the $f(R)$ 
models presented in Ref.~\cite{Hammami2017} (HM). The cyan band represents the 68\% confidence level region, obtained 
using the continuous formation model [Eq.~(\ref{eq:kT1})] and calculated similarly to Ref.~\citep{Afshordi2002} 
(Sec.~3.7). The white dashed line is the average value. 
As expected, deviations from the $\Lambda$CDM model are larger for the model with $f_{R0}=-10^{-4}$, as it represents 
the model with the strongest modifications to gravity. For smaller values of $f_{R0}$, at temperatures $T<1~{\rm keV}$, 
data are in partial agreement with both the $f(R)$ cosmology and the model presented in this work.

Data points have a large dispersion and circumscribe the theoretical models at high mass, while at the lowest masses 
data have a value larger than the simulated HM models and the result of our model. 
Stars show lower masses than the models considered. At high mass, all models are indistinguishable, while at small 
masses differences become visible.

This is because effects of modified gravity depend on the environment and, hence, on the density. In high-density 
regions, screening takes place and deviations from $\Lambda$CDM are smaller. Therefore, in high-density regions the 
$\Lambda$CDM MTR has a similar behavior to that of modified gravity models.
 
Our model shows a non-self-similar behavior and presents a break at $T \simeq 3$~keV. 
At small masses, the slope of the central (average) curve, in the range 0.5-3~keV, is $\simeq 2.3$, and the cyan region 
has an inner and outer slope of $1.8$ and $3$, respectively. 
The quoted bend has been observed in the literature by several authors (see, e.g. \citep{Finoguenov2001}), who, 
assuming the cluster temperature to be constant after the formation time, explained the break as due to the formation 
redshift. 
Another possibility is that the cluster medium is preheated in the early phase of formation \citep{Xu2001}. 
Reference \citep{Afshordi2002}, instead, justified the break with the scatter in the density field. The result of the 
model of Ref.~\citep{Afshordi2002} is shown in Fig.~\ref{fig:comparison}(b) (top right panel), where once again the cyan 
region represents the 68\% confidence level region, (see Ref.~\citep{Afshordi2002}, Sec. 3.7).

This model is not able to distinguish between the effect of formation redshift from scatter in the initial energy of 
the cluster or its initial nonsphericity. 
However, the presence of nonsphericity gives rise to a mass-dependent asymmetric scatter in the MTR. This scatter is 
larger than that of the density field and at small temperatures covers all clusters except one, while the bend in the 
curve of Ref.~\citep{Afshordi2002} takes place almost at the same temperature, $T_X \simeq 3$~keV, in our model. 

In our model, the bend is due to tidal interactions with neighboring clusters, arising from the asphericity of clusters 
(see \citep{DelPopolo1999} for a discussion on the relation between angular momentum acquisition, asphericity, and 
structure formation), and to the effect of dynamical friction. 
Asphericity gives rise to a mass-dependent asymmetric bend in the MTR. The lower the mass, the larger the difference 
from the classical self-similar solution. The origin of the bend is due to a few reasons. Our MTR, differently from 
others (e.g., \citep{Voit2000,Afshordi2002}), contains a mass-dependent angular momentum, $L$, originating from the 
quadrupole moment of the protocluster with the tidal field of the neighboring objects. 
The presence of this additive mass-dependent term breaks the self-similarity of the MTR. 
To be more precise, the collapse in our model is different from the THM: The turnaround epoch and collapse time change, 
as well as the collapse threshold $\delta_{\rm c}$, which is now mass dependent and a monotonic decreasing function of 
the mass (see Fig.~1 in Ref.~\citep{DelPopolo2017}). 
It is larger than the standard value at galactic masses and tends to the standard value when we move to the largest 
clusters. The temperature is $T \propto \epsilon \propto \delta_{\rm c}$ (see \citep{Voit2000}), and then less massive 
clusters are hotter than more massive ones, which are characterized by a standard MTR.

Besides the effect of angular momentum in changing the shape of the MTR, we must recall that another factor 
contributing is the modification of the partition of energy in virial equilibrium, which influences the shape of the 
MT relation. 
At the same time, an important role is played by the cosmological constant and dynamical friction. Both effects, 
similarly to that of angular momentum, delay the collapse of the perturbation. A comparison of the three effects, the 
three terms in Eq. (\ref{eq:coll}), are shown in Fig.~1 of Ref.~\citep{DelPopolo2017} and in Fig.~11 of 
Ref.~\citep{DelPopolo2009}. They are all of the same order of magnitude with differences of a few percent. 
The effect of dynamical friction (DF) was calculated as shown in 
Refs.~\citep{AntonuccioDelogu1994,DelPopolo2006b,DelPopolo2009}.

The first calculations of the role of DF in clusters formation is due to 
Refs.~\citep{White1976,Kashlinsky1984,Kashlinsky1986,Kashlinsky1987}, who considered the DF generated by the galactic 
population on the motion of galaxies themselves. 
Reference \citep{AntonuccioDelogu1994} took into account also the effects of substructure and showed DF produces a 
collapse delay in the collapse of low-$\nu$ peaks, with several consequences, like the mass accumulated by the peak, 
and similarly to tidal torques.

As a consequence of dynamical friction and tidal torques, one expects changes in the threshold of collapse, the 
temperature at a given mass (since $T \propto \delta_{\rm c}$), the mass function, and the correlation function. DF 
and angular momentum have similar effects on structure formation: They delay the collapse, and have similar 
consequences on the collapse threshold.

An important result of the previous calculation is that the MTR in modified gravity cannot be distinguished from that 
predicted by the $\Lambda$CDM model. In HM, the MTR in modified gravity was very different from that of $\Lambda$CDM 
prediction for colder clusters and indistinguishable for hotter ones. Our plots show that the MTR bends in a similar 
way as done by the MTR in the $f(R)$ models and symmetron models (see the following). 
The bending was explained previously, and is related to the effect of several factors as the acquisition of angular 
momentum through tidal torques, by dynamical friction, and by the cosmological constant. 

Our model and the $f(R)$ and symmetron models (see the following) of Ref.~\cite{Hammami2017} are in agreement with data 
till $\simeq 1$~keV; at lower temperatures, a discrepancy is observed with the few clusters present. 
A similar result is found comparing the $f(R)$ models with the model by \citep{Afshordi2002}, in 
Fig.~\ref{fig:comparison}(b). 
In this case, while $f(R)$ models are in disagreement with the data at small masses, this is no longer true for the 
model by Ref.~\citep{Afshordi2002} and $\Lambda$CDM. However, there is a slight disagreement between the model with 
$f_{R0}=-10^{-4}$ and Ref.~\citep{Afshordi2002}.

In particular, in the case of the $f(R)$ models, Figs.~\ref{fig:comparison}(a) and \ref{fig:comparison}(b) show that our 
model is in agreement with all $f(R)$ models considered. In Fig.~\ref{fig:comparison}(b), the slope of the average 
value, in the range 0.5-3~keV, is $\simeq 2.3$, while that of the inner cyan region $\simeq 1.8$ and that of the outer 
cyan region $>3$ for temperatures $<1$~keV.

Fig.~\ref{fig:comparison}(c) shows the same quantities plotted in Figs.~\ref{fig:comparison}(a) and 
\ref{fig:comparison}(b) but for the symmetron case. 
The plot shows that model Sym D is the one which deviates the most from the $\Lambda$CDM, followed by Sym C, Sym B, 
and Sym A. Again, at high mass, till $\simeq 4$~keV, our model, the symmetron models and the data are 
indistinguishable, but Sym D, even if in agreement with the data till $\simeq 3$~keV, slightly differs from our model, 
namely, with the $\Lambda$CDM predictions. 
The discrepancy goes on till $\simeq 2$~keV and then disappears. All the other symmetron models are in agreement with 
our model. As in Fig.~\ref{fig:comparison}(a), for $T \leq 1$~keV, the models are in disagreement with a few clusters. 
Notice that in Figs.~\ref{fig:comparison}(a) and \ref{fig:comparison}(c), we compare the continuous formation model with 
the $f(R)$ and symmetron model, respectively, and then the only change between the two plots is due to the $f(R)$ and 
symmetron curves. The slopes are then the same as in Fig.~\ref{fig:comparison}(a).

Finally, in Fig.~\ref{fig:comparison}(d), we show the same results as in Fig.~\ref{fig:comparison}(c) but for the model 
by Ref.~\citep{Afshordi2002}. The result is similar to Fig.~\ref{fig:comparison}(b). In this case, in the range 
$1 \leq T/{\rm keV} \leq 4$, the model by Ref.~\citep{Afshordi2002} differs from Sym B and D.

The larger discrepancy between the model by Ref.~\citep{Afshordi2002} and the symmetron models in the temperature range 
1-4~keV with respect to the predictions of our model is probably due to the fact that, as stressed by 
Ref.~\citep{Afshordi2002}, the calculation of the effects of the nonspherical shape of the initial protocluster are 
not very rigorous and should be considered as an estimate of the actual corrections. 
The previous assertion is somehow confirmed by the fact that in the given range there is not a real discrepancy 
between cluster data and the other models (except with the model by Ref.~\citep{Afshordi2002}). 

We want to stress that the quoted discrepancies between $\Lambda$CDM predictions and Sym B and D, however, do not imply 
that the symmetron model can be used to claim the MTR is a probe to distinguish between modified gravity and 
$\Lambda$CDM, since in the quoted temperature range there are no visible peculiar differences between the cluster 
data and the model.

As before, we stress that Figs.~\ref{fig:comparison}(d) and \ref{fig:comparison}(b) differ only for the curves relative 
to the $f(R)$ and symmetron models, since we are comparing the last with the same model, namely 
Ref.~\citep{Afshordi2002} [Eq.~(\ref{eq:afsh})]. The slopes are then the same as in Fig.~\ref{fig:comparison}(b).

Finally in Figs.~\ref{fig:comparison1}(a) and \ref{fig:comparison1}(b) we compare the results of the improved top-hat 
model [Eq.~\ref{eq:fin})] with the $f(R)$ [Fig.~\ref{fig:comparison1}(a)] and the symmetron models 
[Fig.~\ref{fig:comparison1}(b)] of Ref.~\citep{Hammami2017}. 
The results are similar to those plotted in Figs.~\ref{fig:comparison}(a) and \ref{fig:comparison}(c), with the 
difference that the slope discussed previously is now smaller. The differences between the model plotted in 
Figs.~\ref{fig:comparison}(a) and \ref{fig:comparison}(c) (revised top hat) and that in Figs.~\ref{fig:comparison1}(a) 
and \ref{fig:comparison1}(b) (continuous formation model) are due to the assumed redshift of formation in the two 
models. 
The slope of the average curve is $\simeq 2$, and those of the outer and inner cyan region, $\simeq 1.8$, and 
$\simeq 2.5$, respectively.

Before concluding, we want to add a note on the redshift dependence of the observed cluster data and the MTR which 
depends on the redshift. All the quantities involved in the determination are, formally, time dependent 
(concentration and temperature). Therefore, when evaluating the MTR, one has to be cautious and aware of this, as the 
time evolution can have a substantial effect on the final result. Nevertheless, in our discussion, redshift evolution 
is not a concern as all the objects considered in Refs.~\cite{Dai2007,Horner1999} are nearby ($z\lesssim 0.2$) and 
neglecting it has a very small impact when compared to the observational error bars on the mass and temperature.

\section{Conclusions}\label{sect:conclusions}
In the present work, we derived the MTR relationship using an improved top-hat model and a continuous formation 
model and compared the results with the prediction of Ref.~\cite{Hammami2017} using $f(R)$ and symmetron models. Our 
model takes into account dynamical friction, the angular momentum acquired through tidal-torque interaction between 
clusters, and a modified version of the virial theorem including an external pressure. The continuous formation model 
is based on the merging-halo formalism by Ref.~\cite{Lacey1993}. 
Both models give a MTR different from the classical self-similar behavior, with a break at 3-4~keV, and a steepening 
with a decreasing cluster temperature. 
The comparison of the quoted MTR with those obtained by Ref.~\cite{Hammami2017} for $f(R)$ gravity and symmetron models 
shows that the MTR is not a good probe to test gravity theories, since the MTR for the $\Lambda$CDM model has the same 
behavior of that obtained by Ref.~\cite{Hammami2017} for the two modified gravity theories considered.

\section*{Acknowledgements}
We thank an anonymous referee whose comments helped us to improve the quality of this work. \\
D.F.M. thanks the Research Council of Norway for their support and the resources provided by UNINETT Sigma2 - the 
National Infrastructure for High Performance Computing and Data Storage in Norway. This paper is based upon work from 
the COST action CA15117 (CANTATA), supported by COST (European Cooperation in Science and Technology). F.P. 
acknowledges support from STFC Grant No. ST/P000649/1.

\bibliographystyle{apsrev4-1}
\bibliography{MTR_MG.bbl}

\end{document}